\documentclass[sigconf,nonacm]{acmart}

\makeatletter
\def\@ACM@bibliography@startnumber{1}
\makeatother
\title{Optimization and Mobile Deployment for Anthropocene Neural Style Transfer}

\author{Po-Hsun Chen}
\orcid{0009-0008-9523-2061}
\affiliation{%
  \institution{Institute of Applied Arts\\National Yang Ming Chiao Tung University}
  \city{Hsinchu City}
  \country{Taiwan (R.O.C)}
}

\author{Ivan C. H. Liu}
\orcid{0000-0002-9226-9765}
\affiliation{%
  \institution{Institute of Applied Arts\\National Yang Ming Chiao Tung University}
  \city{Hsinchu City}
  \country{Taiwan(R.O.C)}
}
\email{ivanliu@nycu.edu.tw}

\begin{document}

\begin{abstract}
This paper presents AnthropoCam, a mobile-based neural style transfer (NST) system optimized for the visual synthesis of Anthropocene environments. Unlike conventional artistic NST, which prioritizes painterly abstraction, stylizing human-altered landscapes demands a careful balance between amplifying material textures and preserving semantic legibility. Industrial infrastructures, waste accumulations, and modified ecosystems contain dense, repetitive patterns that are visually expressive yet highly susceptible to semantic erosion under aggressive style transfer.

To address this challenge, we systematically investigate the impact of NST parameter configurations on the visual translation of Anthropocene textures, including feature layer selection, style and content loss weighting, training stability, and output resolution. Through controlled experiments, we identify an optimal parameter manifold that maximizes stylistic expression while preventing semantic erasure. Our results demonstrate that appropriate combinations of convolutional depth, loss ratios, and resolution scaling enable the faithful transformation of anthropogenic material properties into a coherent visual language.

Building on these findings, we implement a low-latency, feed-forward NST pipeline deployed on mobile devices. The system integrates a React Native frontend with a Flask-based GPU backend, achieving high-resolution inference within 3–5 seconds on general mobile hardware. This enables real-time, in-situ visual intervention at the site of image capture, supporting participatory engagement with Anthropocene landscapes.

By coupling domain-specific NST optimization with mobile deployment, AnthropoCam reframes neural style transfer as a practical and expressive tool for real-time environmental visualization in the Anthropocene.
\end{abstract}

\keywords{Neural Style Transfer, Anthropocene, Mobile Device, Media Arts}

\begin{teaserfigure}
  \includegraphics[width=\textwidth]{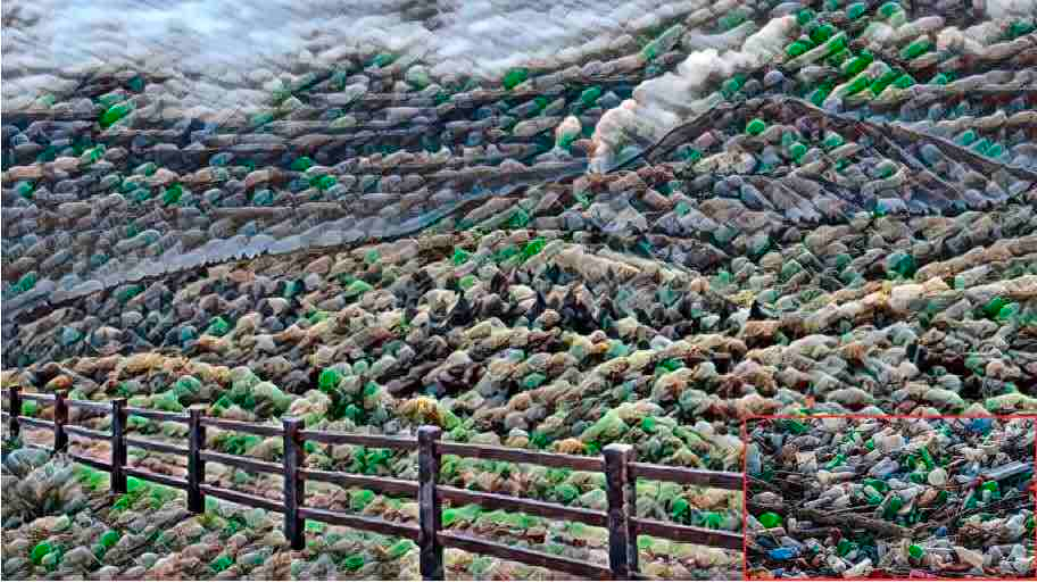}
  \caption{A photo of mountains taken with our mobile system, AnthropoCam, synthesized with an image of plastic waste, shown in the lower-right corner.}
  \label{fig:teaser}
\end{teaserfigure}

\maketitle

\section{Introduction}
The Anthropocene, popularized by Paul Crutzen \cite{crutzen2006anthropocene}, defines a new geological epoch in which human activity has become the dominant force shaping the systems of the Earth. Visually, this era is characterized by industrial signatures such as wastes, modular infrastructure, and modified ecosystems. Although these sites are often socially marginalized or environmentally criticized, they possess a unique aesthetic appeal. This tension is best captured by the concept of the \textit{toxic sublime} \cite{peeples2011toxic} , which describes how media representation can reframe environmental degradation into a space of aesthetic reflection. By representing aesthetics that are typically excluded or suppressed under conventional judgement, we can evoke a sense of beauty at the same time underlining the current environmental issues.

Neural style transfer (NST)\cite{gatys2015neural} is particularly suitable for synthesizing images with Anthropocenic style. Its dense, high-frequency modular patterns, such as cables and pipes, waste accumulations, and concrete buildings, align with NST's ability to encode style through Gram-matrix-based statistics. However, stylizing the Anthropocene presents a unique technical challenge: one must balance the preservation of realism with the expressive amplification of human-altered material traces. Unlike traditional NST research focused on painterly aesthetics, this work investigates the behavior of style transfer when applied to the complex, real-world textures of the human-altered era, Anthropocene.

Drawing on Lev Manovich's \cite{manovich2017instagram} observations on contemporary visual systems and Henry Jenkins' \cite{jenkins2009confronting} framework of participatory culture, we posit that a mobile-based platform is essential. By shifting NST toward mobile accessibility, we enable an immediate, interactive intervention in the surrounding landscape directly at the site of photo taking. This transforms the platform into a vehicle for the public to actively participate in the Anthropocene discourse.

While traditional NST methods are predominantly designed for offline, high-end desktop environments, they lack the responsiveness required for real-time field documentation. We present AnthropoCam, a mobile-based image generation system, enabling mobile accessibility and the stylization of the Anthropocene directly on mobile devices at the site of observation. This mobile accessibility is essential for mass participation of general users or creative enthusiasts. Figure \ref{fig:teaser} shows a photo of mountains taken with our mobile system synthesized with an image of plastic waste. 

Our core contributions are twofold: \textit{Optimized Visual Balancing:} We provide a systematic evaluation of internal NST parameters to achieve a balance between stylistic expression and structural legibility, ensuring the system amplifies Anthropocene textures without compromising its integrity. \textit{Low-Latency Mobile Pipeline:} A React Native and Flask architecture utilizing feed-forward \cite{johnson2016perceptual} networks to achieve 3–5s high-resolution inference on general mobile hardware. 

By situating technical optimization within a mobile context, AnthropoCam transforms NST into an aesthetic research interface for reinterpreting the more-than-human in the Anthropocene.

\section{Related Works}
\subsection{Neural Style Transfer and Perceptual Loss}

Neural style transfer (NST) was introduced by Gatys et al. \cite{gatys2015neural}, who demonstrated that deep convolutional neural network (CNN) pretrained for object recognition implicitly encode separable representations of image content and style. By matching Gram matrix statistics of feature maps, their method enables the recombination of content structure and stylistic appearance. This formulation established NST as a technique grounded in texture statistics rather than specific semantic synthesis.

Subsequent work by Johnson et al. \cite{johnson2016perceptual} introduced perceptual loss functions to train feed-forward networks for real-time style transfer, significantly reducing inference time while preserving visual quality. Dumoulin et al. \cite{dumoulin2016learned} further proposed conditional instance normalization to support multiple styles within a single network, enabling more flexible style representation---particulariy suitable for deployment on mobile devices. These approaches emphasize controllability, making NST suitable for domain-specific style exploration rather than purely generative tasks.(Fig 2)

Unlike GAN or diffusion-based image synthesis, NST maintains a direct correspondence to the input content image, ensuring fidelity while transforming appearance. This property is particularly important when we are performing our conceptual style transfer, applying a specific style to the image without destroying the original image, because authenticity and recognizability remain crucial.

\begin{figure}[h]
  \centering
  \includegraphics[width=\linewidth]{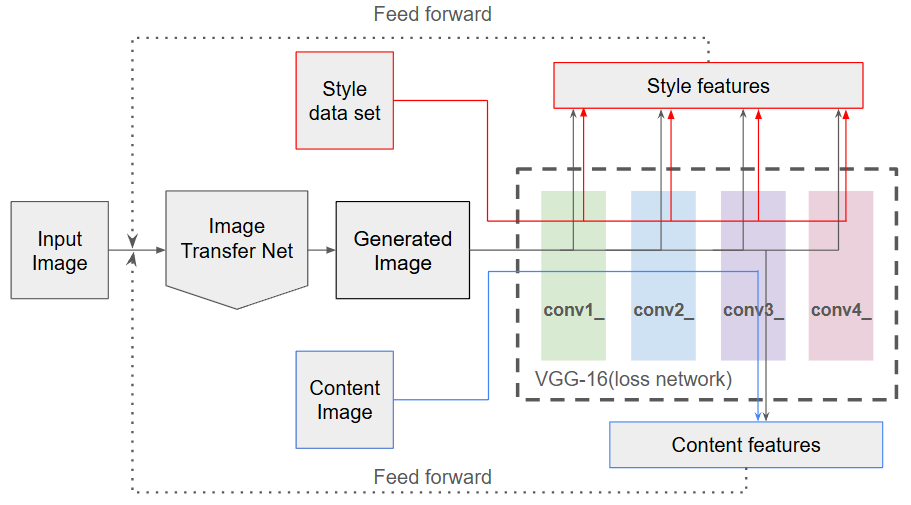}
  \caption{Overall system architecture for neural style transfer: Image Transform Net is the image style transfer system, belonging to the feed-forward computation system; Loss network is the loss calculation system, and the calculated loss is backpropagated to update the Image Transform Net.}
  \label{fig:archi}
\end{figure}

\section{System Overview}

\begin{figure}[h]
  \centering
  \includegraphics[width=\linewidth] {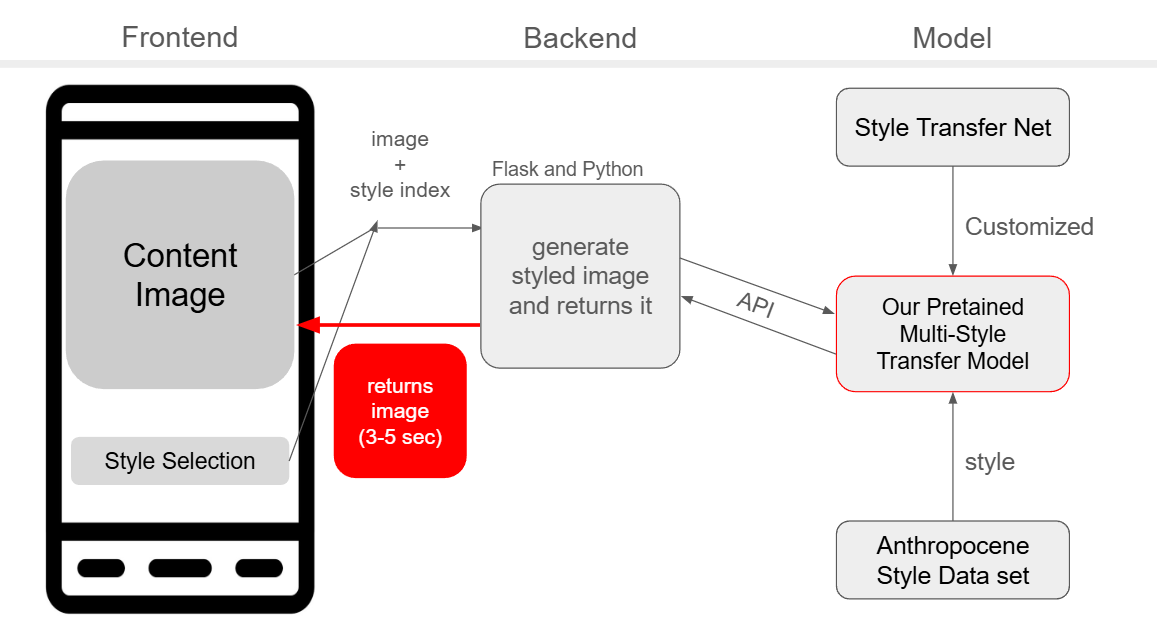}
  \caption{System Overview: This system combines a front-end application interface (mobile device) with content and style selection, and a pre-trained style transfer model on the back-end to achieve near real-time style transfer on mobile devices.}
  \label{fig:system}
\end{figure}

AnthropoCam is optimized for the constraints of mobile hardware. By bridging high-performance backend inference with reactive front-end, we address the latency and resolution trade-offs inherent to real-time environmental documentation. As illustrated in Figure \ref{fig:system}, the system integrates three core components:

\textit{Mobile Interface (Front-end):} This layer handles image capture and style selection, transmitting the content and style picking to the server.

\textit{Inference Pipeline (Backend):} A Python server executes a single model evaluation, optimized to return high-resolution results within 3–5 seconds.

\textit{Stylization Network (Models):} We utilize pre-trained network with Anthropocene images, based on the method proposed by Dumoulin et al. \cite{dumoulin2016learned}, enabling instantaneous transformation without iterative optimization.

By bridging offline parameter experimentation with mobile-based image generation, the system provides a platform to evaluate the technical feasibility and aesthetic appeal of the Anthropocene style transfer.

\section{Method: Neural Style Transfer and Mobile Deployment}

Our framework adapts the NST formulation to the specific visual domain of the Anthropocene. The system utilizes a pre-trained \cite{dumoulin2016learned} VGG-16 network as the backbone for feature extraction and perceptual loss calculation.

\subsection{Feature Representation and the Gram Matrix}

The framework utilizes a convolutional neural network (CNN) as the primary model for feature extraction, following the representation established by Gatys et al. \cite{gatys2015neural}.
Let $F^l \in \mathbb{R}^{N_l \times M_l}$ be the feature activation at layer $l$ of the VGG-16 network. Within this architecture, the numerical indices of the layers are proportional to their depth. A larger $l$ implies a deeper layer in the network. Specifically, we denote layers using the ``conv'' prefix followed by two identifiers (e.g., conv4\_2): the first digit represents the sequential block within the network, while the second digit, following the underscore, refers to the individual sub-layer within that specific block.

In the experimentation throughout this paper, the feature extraction of the content image was fixed at $l=$conv3\_3 as the controlled parameter. This specific layer was chosen for balanced visual details while keeping semantic structure. The style image was characterized via the Gram Matrix $G^l \in \mathbb{R}^{N_l \times N_l}$, which represents correlations between feature maps $F_{ij}^l$-cross specific CNN layers $l$:
\begin{equation}
  G_{ij}^l = \sum_k F_{ik}^l F_{jk}^l
  \label{eq:Grammatrix}
\end{equation}

In the current context of Anthropocene style transfer, this matrix effectively encodes the repetitive patterns of industrial materials (e.g., pipes or cables) across layers conv1\_2, conv2\_2, conv3\_3, and conv4\_3.

\subsection{Multi-Objective Loss Function}

To optimize the generated image $x$, we minimize a joint loss function that balances the structural integrity of the content image $p$ against the stylistic expression of target $a$. This optimization is governed by the weighted sum of content $\alpha$, style $\beta$, and total variation (TV) $\gamma$ losses, described below:

\textit{Content Loss} ($\mathcal{L}_{content}$):
Calculates the Euclidean distance between the feature maps of the content image $p$ and the generated image $x$:
\begin{equation}
  \mathcal{L}_{content}(p, x, l) = \frac{1}{2} \sum_{i,j} (F_{ij}^l(x) - F_{ij}^l(p))^2
  \label{eq:CLoss}
\end{equation}

\textit{Style Loss} ($\mathcal{L}_{style}$):
Measures the discrepancy in texture statistics between the style target $a$ and $x$ across multiple layers, where $w_l$ is the weighting factor for each layer. Our experiments specifically test the impact of these weights on Anthropocene textures.

\begin{equation}
    \mathcal{L}_{style}(a, x) = \sum_l w_l \frac{1}{4N_l^2 M_l^2} \sum_{i,j} (G_{ij}^l(x) - G_{ij}^l(a))^2
    \label{eq:SLoss}
\end{equation}

\textit{Total Variation Loss} ($\mathcal{L}_{tv}$):
To ensure visual smoothness and reduce high-frequency artifacts (noise) common in mobile-captured images, we incorporate a Total Variation regularizer:
\begin{equation}
    \mathcal{L}_{tv}(x) = \sum_{i,j} \left( (x_{i,j+1}-x_{i,j})^2 + (x_{i+1,j}-x_{i,j})^2 \right)^{1/2}
    \label{eq:TVLoss}
\end{equation}

\subsection{Total Loss and Optimization}

The final loss function is:

\begin{equation}
    \mathcal{L}_{total} = \alpha \mathcal{L}_{content} + \beta \mathcal{L}_{style} + \gamma \mathcal{L}_{tv}
    \label{eq:totalLoss}
\end{equation}
where $\alpha/\beta$ ratio and $\gamma$ are the primary control variables in our experimental analysis. Unlike iterative optimization \cite{gatys2015neural}, we implement a feed-forward style transfer network \cite{johnson2016perceptual} for our mobile deployment, enabling near real-time inference ($<$ 5s) by mapping the optimization problem into a single forward pass.

\subsection{Implementation and Inference Pipeline}

To satisfy the low-latency requirements of mobile interaction, we depart from traditional iterative optimization \cite{gatys2015neural} methods in favor of a feed-forward image-transformation architecture \cite{johnson2016perceptual}. Once trained on a specific Anthropocene style, the network performs stylization in a single forward pass without further iterations, ensuring near real-time results. Our system stack leverages a Flask-based backend on a GPU-enabled server to handle inference requests, while the React Native frontend facilitates cross platform image capture and preprocessing, images are normalized and resized to a consistent scale before inference, with the final generated result transmitted back via an API to maintain a seamless user experience.

\section{Experiments on Optimization Parameters}

This section evaluates the AnthropoCam framework through a series of controlled experiments designed to analyze the NST parameters and the resulting visual characteristics of the Anthropocene.
We begin from fundamental feature representation, which focuses on style synthesis, we determine how layer selection, style dataset and epoch counts influence the model's capacity, the goal is to ensure that the expressive amplification of industrial textures does not lead to semantic erasure, but instead maintains a balance between stylistic strength and structural integrity.

Then we shifts toward system-level performance optimization, focusing on the framework's viability in a real-world mobile context. We analyze the trade-offs between the resolution and latency to establish the optimal configuration for mobile hardware. By correlating stylistic texture with computational speed, these experiments ensures the system delivers acceptable results within 3-5 seconds.

\subsection{Feature Representation and Style Compatibility}

The efficacy of Anthropocene style transfer is fundamentally governed by the hierarchical feature extraction layers of the VGG-16 backbone. In this section, we analyze how layer selection and the statistical composition of the style dataset influence the visual translation of anthropogenic textures.
\subsubsection{Layer Selection for the Anthropocene Textures}

\begin{figure}[h]
  \centering
      \includegraphics[width=\linewidth]{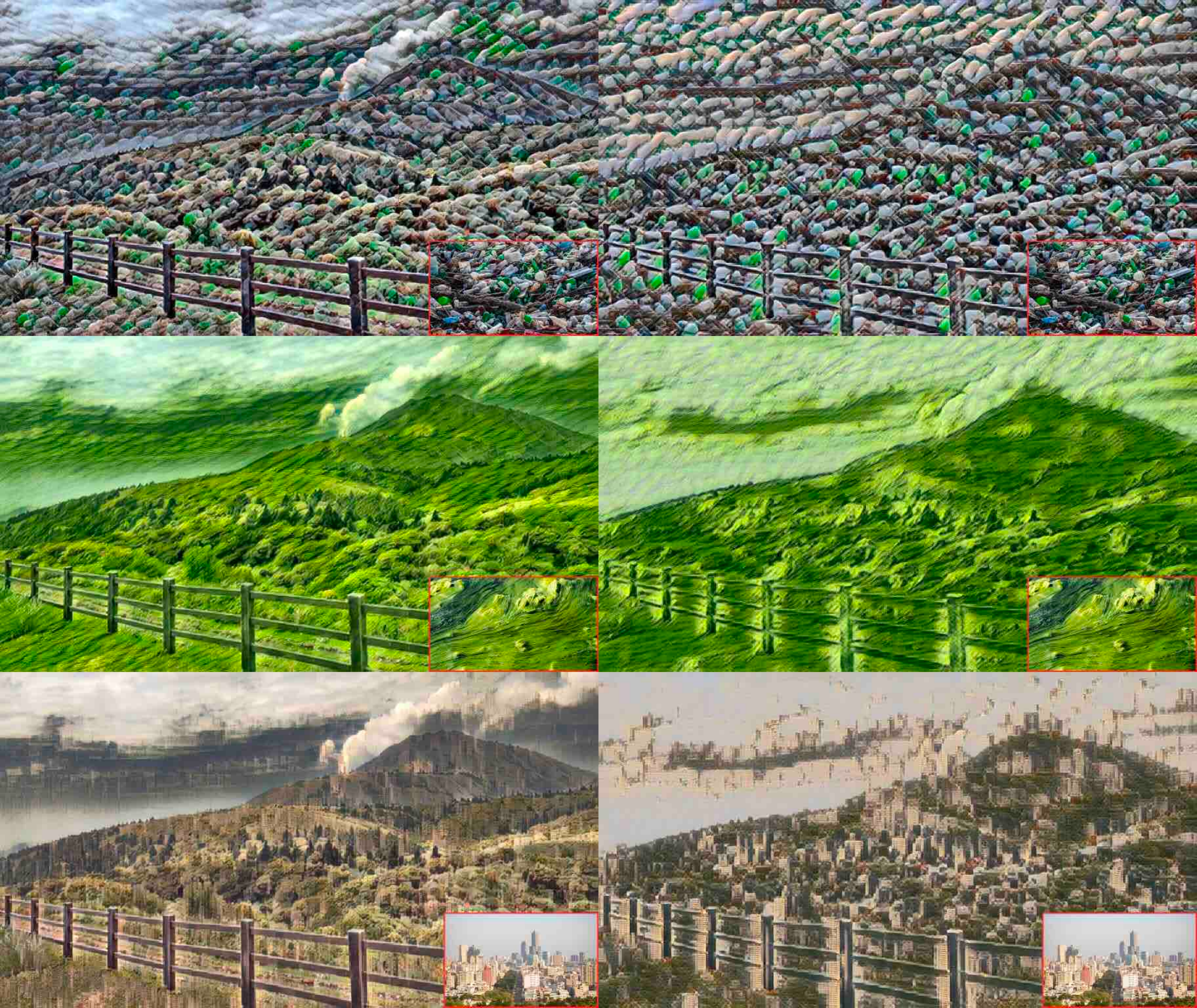}
  \caption{The left column shows features from shallower layers with more detailed textures. The right column shows features from deeper layers, resulting in a more blocky image. (From top to bottom: plastic bottle, eutrophication, and urban building style)}
  \label{fig:layer}
\end{figure}

The selection of content and style layers in NST strongly influences the trade-off between structural fidelity and stylistic abstraction. Shallower style layers emphasize fine-grained textures and high-frequency details, while deeper layers reveal more spatial correlation resulting in block-like stylization, Equations 2 \& 3. We found that style images with different sizes of visual details tend to prefer different depths of layers. There are two observed findings:

\textit{Modular Infrastructures} (e.g., shipping containers): These subjects are characterized by rigid geometries and large-scale structural repetitions. We found that utilizing deeper style layers (e.g., $l$ = conv4\_2, conv4\_3) is essential for capturing these block-like abstractions. The larger receptive fields in these layers allow the model to ignore fine-grained noise and focus on the spatial modularity of the industrial form, see Figure \ref{fig:layer} (left column).

\textit{Filamentous Patterns} (e.g., eutrophication patterns): These subjects rely on high-frequency textural details. For such styles, we prioritize shallower layers (e.g., $l$ = conv2\_2, conv3\_1). These layers preserve the micro-features, such as the sharp edges of the organic filaments in eutrophicated water, which are otherwise lost in the spatial downsampling of deeper layers, see Figure \ref{fig:layer} (right column).

\subsubsection{Style Dataset Interference and Homogeneity}

\begin{figure}[h]
  \centering
      \includegraphics[width=\linewidth]{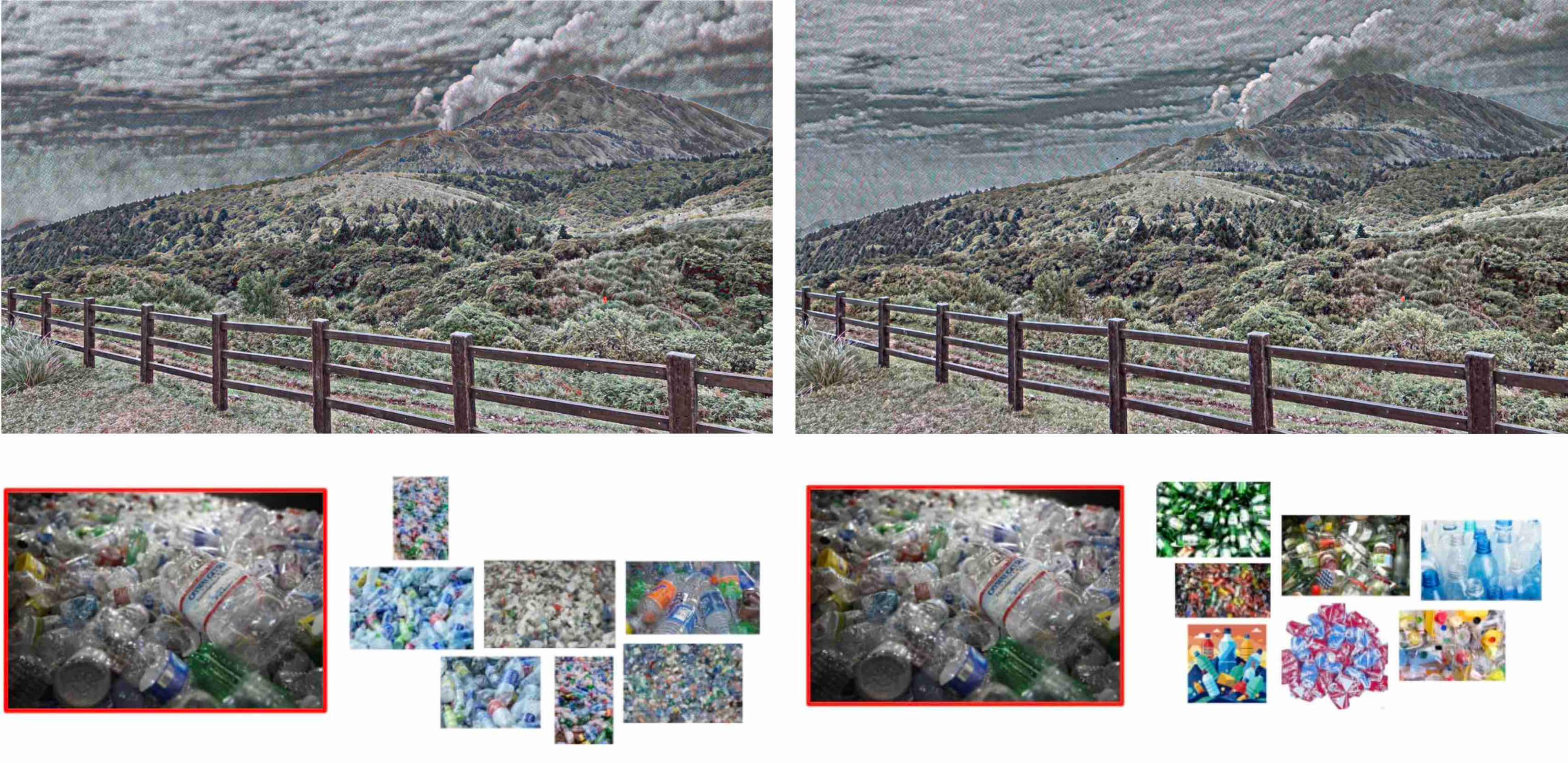}
  \caption{Effect of training set consistency on resulting visual textures. The left set of images show the output using more visually consistent style images, while the right uses more contrasting style images. The bottom row shows the style image sets used during training. It can be seen that similar styles maximize the presentation of style features (notice also the difference in clouds).}
  \label{fig:dataset}
\end{figure}

A critical finding in our training pipeline is the impact of different style images in the dataset on the convergence stability, style can be influenced by other image features from the style data set, see Figure \ref{fig:dataset}.

Through a comparative analysis, we identified a phenomenon of inter-style interference: when training with a set of style images that are highly different in contrast or colors, we observed a dilution of visual characteristics. As in the right panel of Figure \ref{fig:dataset}, we found that although the model remains functional, the result often lacks the strength of stylization, becoming weaker due to the averaging effect of conflicting color statistics.

In contrast, the left panel of Figure \ref{fig:dataset} demonstrates the advantages of utilizing a visually consistent training set. When the style images maintain a consistency, it results in a more texturally pronounced output. Our findings suggest that maintaining such consistency leads to a more assertive and coherent visual output, accurately capturing the complex texture of the Anthropocene.

\subsubsection{Transparency and Localized Style Selection}

Beyond shape and texture, we identified two procedures that significantly enhance the aesthetic appeal of the Anthropocene.

\begin{figure}[h]
  \centering
      \includegraphics[width=\linewidth]{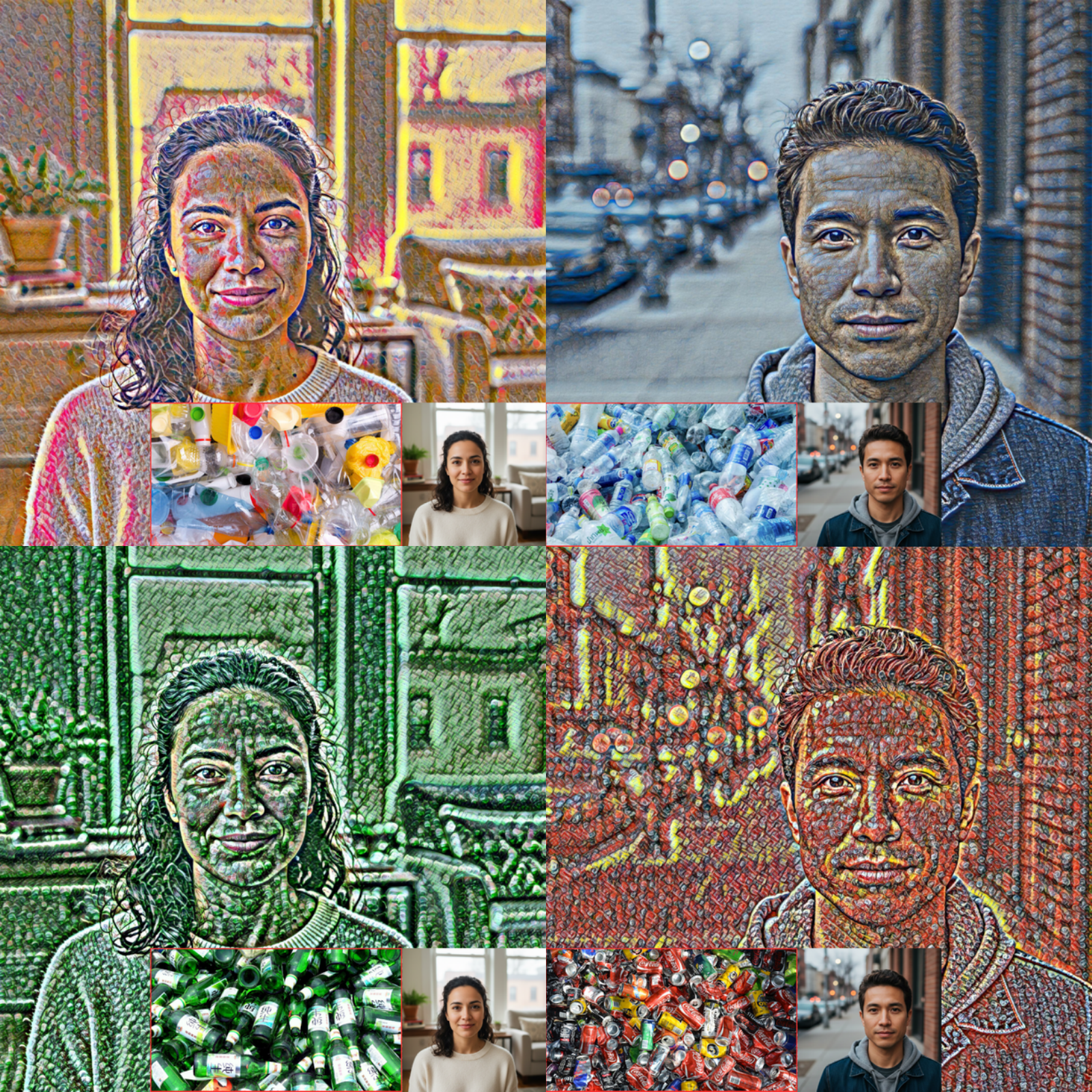}
  \caption{Images in the top row feature semi-transparent objects, resulting in a hazy, ethereal quality, compared to the more vibrant colors in the bottom row.}
  \label{fig:human}
\end{figure}

\textit{Transparency}: Figure \ref{fig:human} shows AI-generated (non-human) portrait images that used style images with translucent or semi-transparent regions (e.g., piles of plastic) introduces a hazy effect in the generated output, this results in a soft-focus diffusion of the content edges.

\begin{figure}[h]
  \centering
      \includegraphics[width=\linewidth]{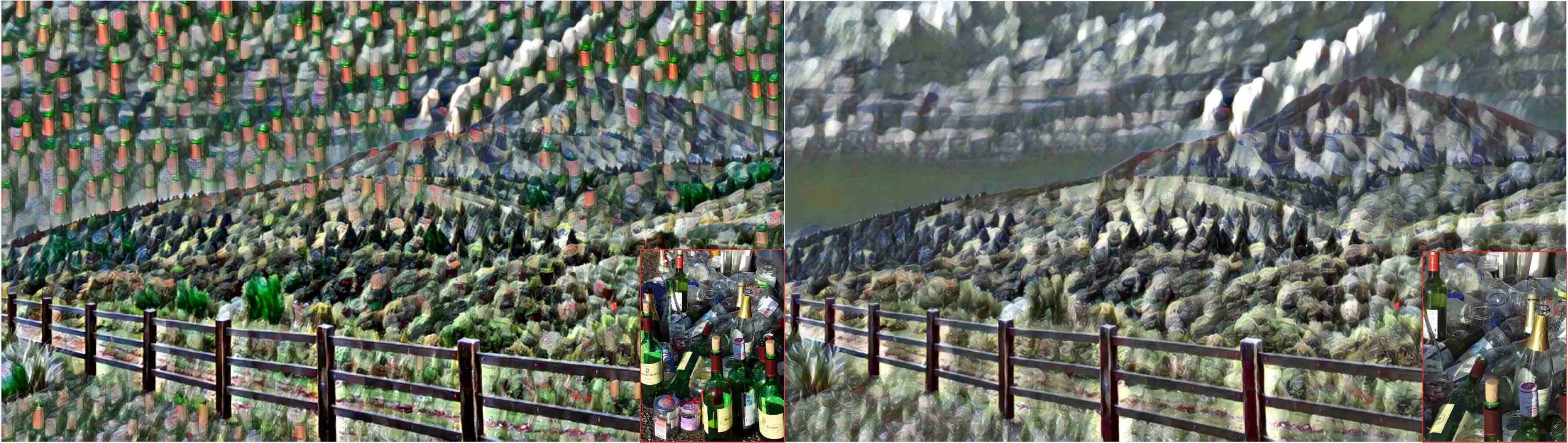}
  \caption{Using the same style images for local testing, with the original image on the left and the local image on the right, the features will be significantly different.}
  \label{fig:local}
\end{figure}

\textit{Localized Style Sampling}: To balance the preservation of original content with stylistic expression, we utilized localized sampling of the style source. Figure \ref{fig:local} reveals that by cropping specific regions of a style image that contain representative textures (e.g., glass bottles without labels), we can effectively constrain the color and shape transformation. This prevents the stylization from overwhelming the semantic structure of the scene while maintaining the texture of the original Anthropocene artifact.

\subsection{Balancing Style and Semantic Legibility}

The optimization of Anthropocene style transfer involves a delicate trade-off between stylistic strength and the preservation of the original content's structural integrity. This section details our empirical findings regarding loss weighting, training epochs, and the stochastic dynamics of the optimization process.

\subsubsection{Style Weight Sensitivity and the Legibility Threshold}

\begin{figure}[h]
  \centering
      \includegraphics[width=\linewidth]{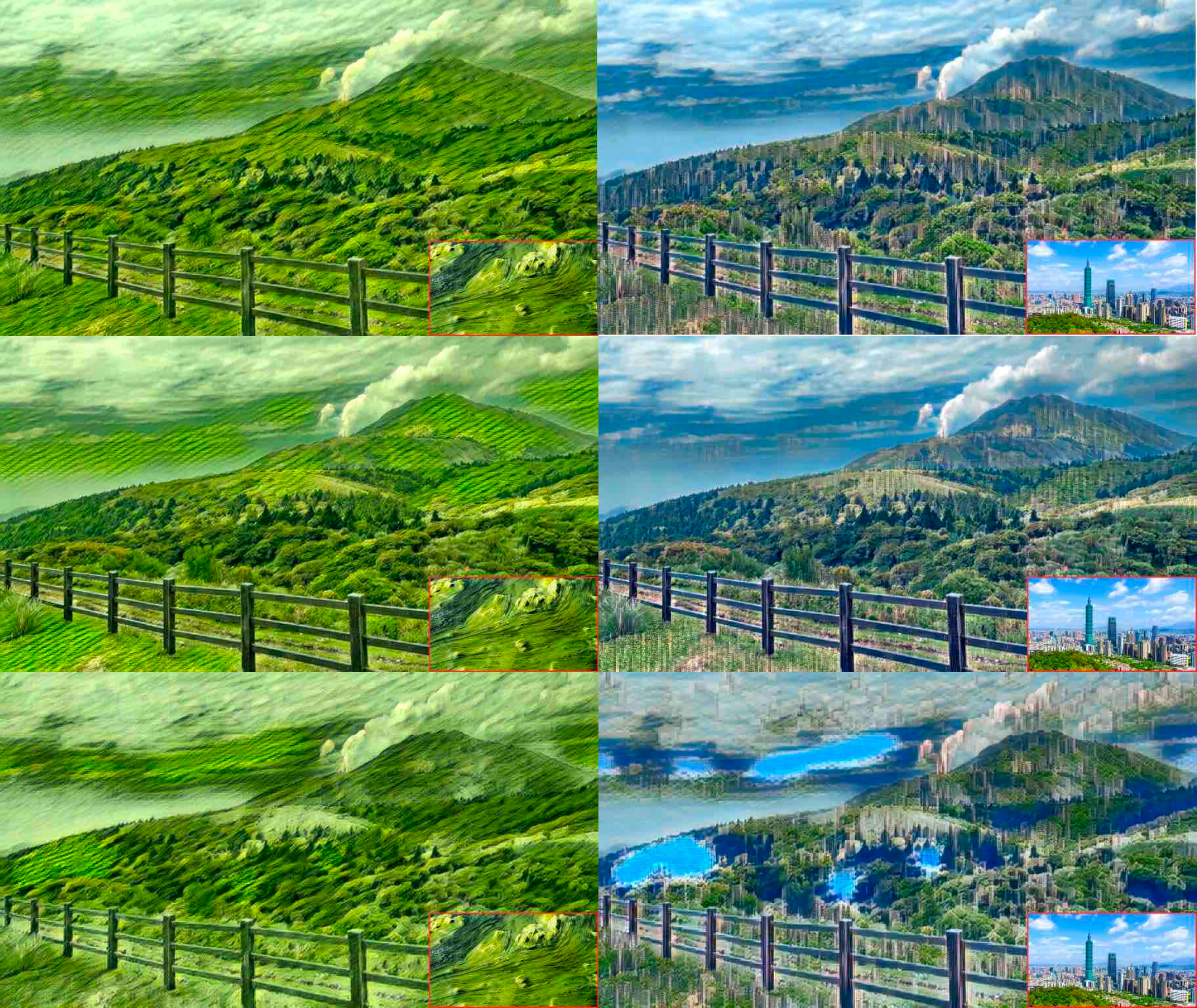}
  \caption{The weight of style in the total loss is controlled by the following values, from top to bottom: $w_s = 2$, $5$, and $8$. Higher values result in a more pronounced style effect, but excessively high values can lead to content distortion and loss of quality.}
  \label{fig:styleweight}
\end{figure}

To ensure the fidelity of content images, Figure \ref{fig:styleweight} fixed the content weight $w_c = 1$ and systematically varied the style weight, ranging from $w_s = 2$, $5$ to $8$ in Equation \ref{eq:SLoss}. We analyze the results below:

\textit{Subtle} ($w_s = 2$): Top row of Figure \ref{fig:styleweight} shows that, at this lower bound, the generated images maintain high structural stability. However, the stylistic translation is insufficient to deliver visual impact required.

\textit{Optimal balance} ($w_s = 5$): The second row of Figure \ref{fig:styleweight} shows the most balanced result. It enables sufficient style expression, capturing color palettes and textures, while keeping the total loss low enough for semantic recognition.

\textit{Semantic erasure} ($w_s = 8$): Bottom row of Figure \ref{fig:styleweight} shows that excessive weight leads to a surge in total loss, resulting in blocky artifacts and high-frequency noise that obscure the content's underlying geometry.

\subsubsection{Temporal Convergence and Perceptual Saturation}

\begin{figure}[h]
  \centering
  \includegraphics[width=\linewidth]{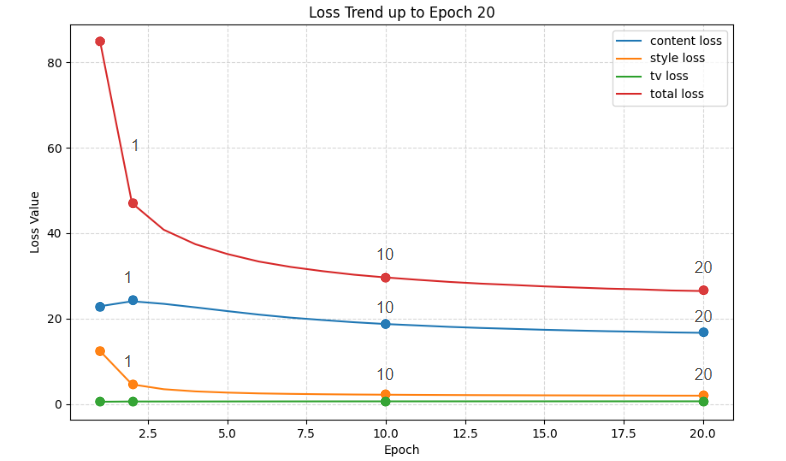}
  \caption{Loss convergence status upto 20 epochs. Initial convergence shows a significant reduction in loss. However, the imporvement on convergence is negligible between epochs 10 and 20. }
  \label{fig:epoch}
\end{figure}

In NST, a training \textit{epoch} signifies one complete pass through the dataset, consisting of a series of iterations determined by the batch size. In Figure \ref{fig:epoch}, we evaluated the final epoch counts by comparing loss results between 1, 10, and 20 epochs. While training with a few iterations significantly reduces computational time, the quality severely lacks in style depth and detail compared to a single training epoch, the quality of feature learning depends heavily on epoch counts:

\textit{1 epoch}: Achieved better convergence to thousands of iterations, but failed to reproduce the characteristics of the Anthropocene style.

\textit{10 epochs}: Increasing from 1 to 10 epochs yielded significant improvements in the model's ability to learn complex textures, such as the specific brushstrokes and shapes.

\textit{20 epochs}: Beyond 10 epochs, we observed the saturation of visual effect improvement. Although the training loss continues to decrease, see in Figure \ref{fig:epoch}, the convergence of the loss and the visual improvements are marginal. Furthermore, the model risks overfitting to specific noise patterns within the style dataset, which can degrade the generalization performance on diverse mobile-captured content.
\subsubsection{Optimization Stability and Visual Smoothing}

\begin{figure}[h]
  \centering
      \includegraphics[width=\linewidth]{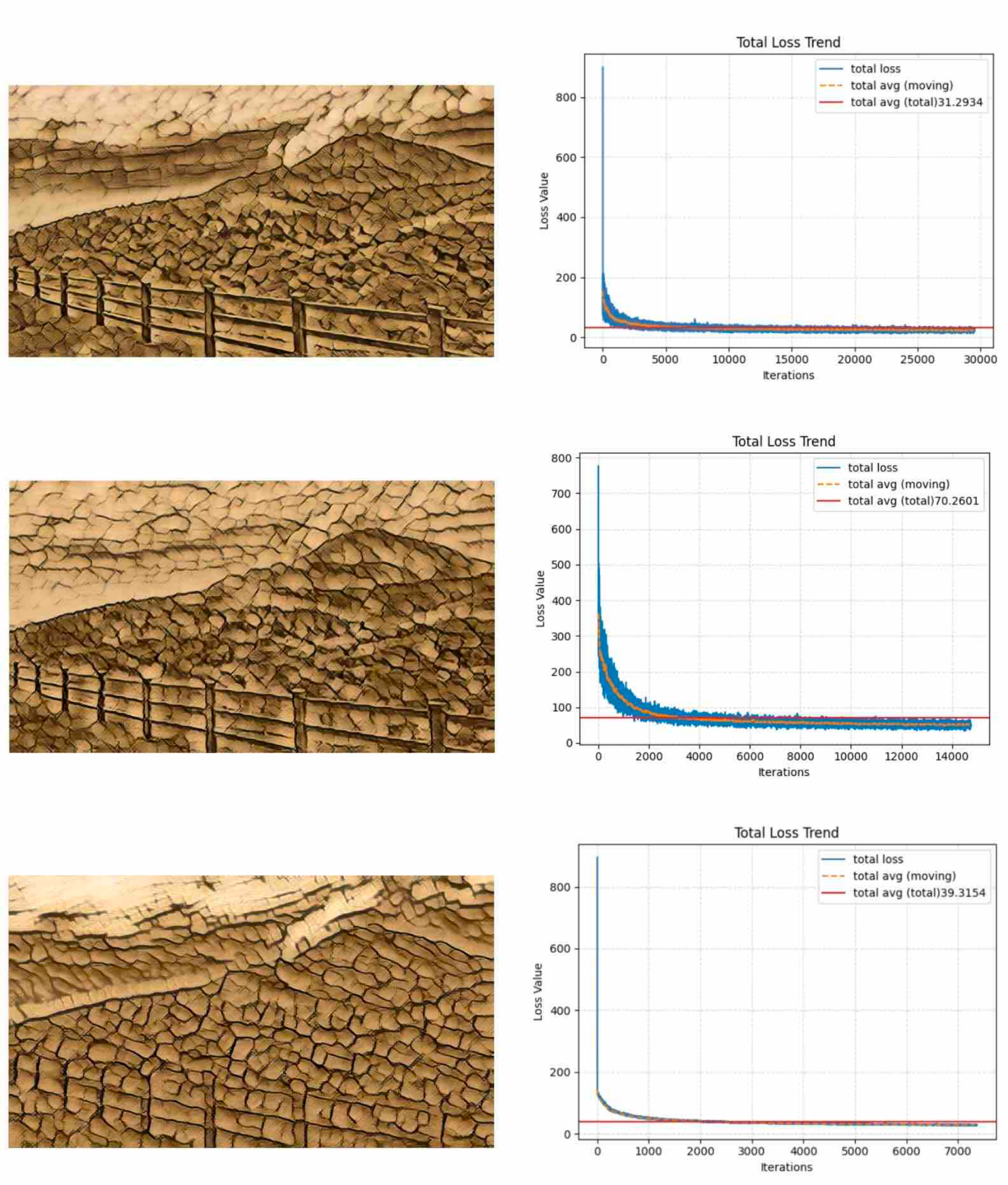}
  \caption{This figure illustrates the convergence during training, from top to bottom shows batch size of $n=4$ ,$n=8$  and $n=16$. As observed in the loss curves(right row), batch size of $n=4$ resulting in more pronounced fluctuations during the update process, while batch size of $n=16$ results in higher memory cost and longer time.}
  \label{fig:batch}
\end{figure}

We evaluated batch sizes between $n=4$, $8$ and $16$ to balance stability and efficiency. In Figure \ref{fig:batch}, small batches ($n=4$) caused oscillatory convergence and large batches ($n=16$) increased computational overhead, $n=8$ provided the optimal trade-off. Apart from memory efficiency, smaller batch sizes introduces greater variation of surface gradient in the optimization landscape, preventing the loss function from becoming trapped in local minima and ensuring stable convergence.

To mitigate high-frequency noise and boundary fragmentation often induced by dense industrial textures, we integrated total variation (TV) Loss, Equation \ref{eq:TVLoss}, as a noise reduction mechanism. By penalizing abrupt pixel-wise gradients, the TV loss in Equation \ref{eq:totalLoss} maintains visual continuity and prevents shattered artifacts, ensuring that the final output preserves the materiality of the Anthropocene subject even under strong stylization.
\subsection{Effect of Output Resolution on Efficiency}

The deployment of AnthropoCam on mobile devices necessitates a strategic balance between stylistic textures and inference latency. Our experiments reveal that output resolution is not merely a performance variable but a critical factor influencing the visual effect of the stylization.

\subsubsection{Visual Texture vs. Resolution}

\begin{figure}[h]
  \centering
      \includegraphics[width=\linewidth]{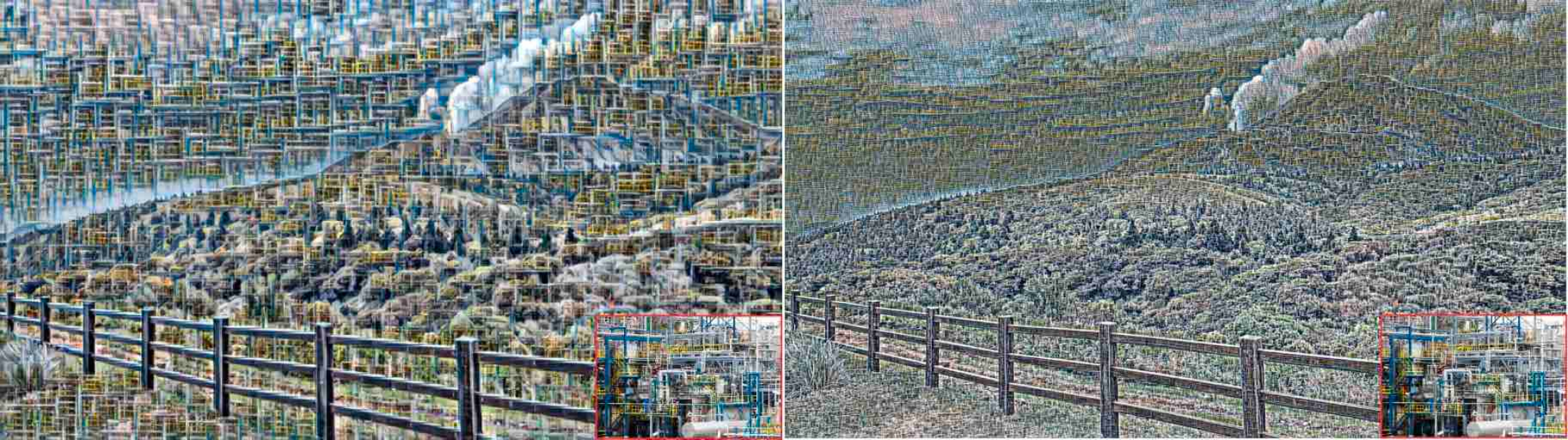}
  \caption{Stylistic texture vs. Resolution. (Left) Lower-resolution: Larger relative receptive fields result in modular abstraction, simplifying textures into geometric blocks. (Right) Higher-resolution: Smaller relative coverage preserves high-frequency details, capturing the dense repetitions of the Anthropocene.}
  \label{fig:pixel}
\end{figure}

Anthropocene textures varies significantly across different output resolutions:

\textit{High-resolution} (1920 $\times$ 3416px): In Figure \ref{fig:pixel} (left), the receptive field of the convolutional filters covers a smaller relative area of the image. This results in finer, high-frequency textural details and dense pattern repetitions, which are ideal for capturing the intricate weathering of industrial textures. However, this comes at the cost of significantly higher computational overhead.

\textit{Low-resolution} (540 $\times$ 960px): In Figure \ref{fig:pixel} (right), the same filters cover a larger proportion of the content image, leading to a more modular, block-like abstraction. Although this maximizes inference speed, the textural complexity of the synthesized image is often oversimplified into coarse geometric shapes.

\subsubsection{Optimization for Near Real-Time Feedback}
Considering the scenario for usage on mobile devices, we set a tolerance on the processing latency to be 3-5 seconds. To ensure a responsive mobile experience, we determine the optimal operational resolution based on inference latency and visual quality: while high-resolution processing (1920 $\times$ 3416px) successfully captures intricate details, its latency exceeds the tolerance threshold for interactive mobile use. Conversely, although low resolutions (540 $\times$ 960px) offer instant feedback, the resulting blocky effect fails to convey the details of the Anthropocene aesthetics. We identified (1280px $\times$ 2276px) as the optimal compromise for our mobile pipeline. At this scale, the system preserves sufficient textural detail and structural complexity while maintaining an acceptable latency on general mobile hardware, enabling users to take photos of their surroundings and synthesize without sacrificing the textural details.

\section{Discussion and conclusion}

This research presents AnthropoCam, a mobile-based neural style transfer framework tailored for capturing and transforming Anthropocene landscapes. By systematically optimizing model parameters, we achieved a critical balance between stylistic expression and structural legibility, addressing the common challenge of semantic erasure in industrial texture synthesis. Our findings demonstrate that the strategic selection of \textit{conv} layers and style weights allows the system to amplify the texture of human-altered environments without compromising the environmental context essential for real-time photography. Furthermore, the integration of total variation (TV) loss ensures visual smoothing and gradient stability, preventing high-frequency noise from disrupting the user’s real-time style transfer process.

On a practical level, the implementation of a feed-forward architecture proves highly effective for mobile applications. Unlike traditional iterative methods, AnthropoCam delivers high-resolution results within a 3–5 second, meeting the low latency demands of real-time image capture. This integration of a Flask-based GPU backend and a React Native frontend illustrates how sophisticated deep learning machine can be applied into accessible general mobile tools, fostering new aesthetic perspectives on environmental change.

In conclusion, this work establishes a comprehensive pipeline from fundamental feature representation to system-level optimization. Moving forward, the framework will evolve through a data-driven optimization loop: by capturing anonymized user behavior and style preferences, the system will gain the capacity to identify emerging aesthetic trends. These collective insights will drive automated model updates and iterations, allowing the system to refine its outputs and parameters based on real-world usage and ensuring that the visual representation of the Anthropocene continues to evolve alongside collective human perception.

\section*{Acknowledgements}
We thank Wan-Tien Wu for helpful discussions, and Ya-Chen Hsu for earlier investigation on neural style stransfer.

\bibliographystyle{abbrv}
\bibliography{ref}

\end{document}